\documentclass[a4paper]{jpconf}
\usepackage{graphicx}
\usepackage{amsmath,amsthm,latexsym,amssymb,amsfonts,epsfig, psfrag,url}
\usepackage{dsfont}
\def\be{\begin{equation}}
\def\ee{\end{equation}}
\def\ba{\begin{eqnarray}}
\def\ea{\end{eqnarray}}
\def\q{{\quad}}

\begin{document}
\title{Effective relational dynamics\footnote{Based on a talk at the conference {\it Loops '11} in Madrid on May 24th 2011 \cite{talk}.}}

\author{Philipp A H\"ohn}

\address{Institute for Theoretical Physics, Universiteit Utrecht, 3508 TD Utrecht, The Netherlands}

\ead{p.a.hohn@uu.nl}

\begin{abstract}
We provide a synopsis of an effective approach to the problem of time in the semiclassical regime. The essential features of this new approach to evaluating relational quantum dynamics in constrained systems are illustrated by means of a simple toy model. 
\end{abstract}

\section{The problem of time}

The application of the Dirac quantization programme to generally covariant systems triggers a `conceptual problem' since the quantization of the Hamiltonian constraint yields a seemingly `timeless' evolution equation $\hat{H}|\psi\rangle=i\hbar\frac{\partial}{\partial t}|\psi\rangle=0$ and thereby leads to the issue of `frozen dynamics' in the quantum theory. While classically the presence of a Hamiltonian constraint does not lead to an impeding obstacle because we can always resort to a time coordinate along the flow of the Hamiltonian with respect to which one can order physical relations, such a time coordinate is absent in the physical Hilbert space. The standard `conceptual solution' to this problem is {\it relational dynamics} \cite{rel}, according to which time evolution arises by relating dynamical quantities to other internal variables which we use as clocks to keep track of internal time; such relational statements are invariant under the gauge flow of the Hamiltonian constraint and could thus, in principle, be promoted into well--defined operators on the physical Hilbert space. Nevertheless, even when adopting this `conceptual solution' one encounters a whole plethora of {\it technical} problems in the quantum theory \cite{rel,pot}. Here we want to address four such problems:
\begin{enumerate}
\item The {\it Hilbert space problem}. Which Hilbert space representation is one to use and, in particular, how is one to construct a positive-definite inner product on the space of solutions?
\item The {\it multiple choice problem}. Which relational clock variable should one employ? (Different relational clocks may yield {\it a priori} different quantum theories.)
\item The {\it global time problem}. Good global clocks which parametrize the gauge orbits such that every trajectory intersects every constant clock time slice once and only once may not exist.
\item The {\it observable problem}. It is a notoriously difficult challenge to construct explicit observables in generally covariant theories, especially in the quantum theory.
\end{enumerate} 
In the sequel we summarise the effective approach to the problem of time \cite{bht1,bht2} which---at least in the semiclassical regime---deals with some of these problems and circumvents others. Rather than employing special clock choices and attempting to justify these, we follow the basic premise that there are to be no distinguished clocks and that, instead, we ought to treat all clocks on an equal footing. Our goal consists in employing local internal times, possibly translating between different clocks and thereby evolving (relational) data along complete semiclassical trajectories.

\section{Effective description of constrained systems}

The central idea of the effective approach is to sidestep the {\it Hilbert space problem} altogether; instead of employing wave functions or density matrices to describe states in a fixed Hilbert space, we regard states as ({\it a priori} complex) linear functionals on an algebra of kinematical variables, say polynomials in a canonial pair $(q,p)$, and use expectation values $\langle\hat{q}\rangle$
and $\langle\hat{p}\rangle$, and moments\footnote{The subscript `Weyl' stands for Weyl--ordering, i.e.\ total symmetrization of the operators inside $\langle\ldots\rangle$.} 
\[
 \Delta(q^{a}p^{b} ):=\langle(\hat{q}-\langle\hat{q}\rangle)^{a}
(\hat{p}-\langle\hat{p}\rangle)^{b} \rangle_{\rm Weyl},\,\,\,\,\,\, \,\,\,\,\,\,\,\,\,\,\,\, a+b\geq2
\]
to parametrize states \cite{bojski}. This construction immediately carries over to an arbitrary number of canonical pairs. 

This space of states can be given a (quantum) phase space structure via a Poisson bracket defined as follows for any operators $\hat{A},\hat{B}$ polynomial in the canonical variables\footnote{This definition of a Poisson bracket on quantum phase space can be motivated by the geometrical formulation of quantum mechanics \cite{bojski}.}
\ba\label{poisson}
 \{\langle\hat{A}\rangle,\langle\hat{B}\rangle\}:=
\frac{\langle[\hat{A},\hat{B}]\rangle}{i\hbar}.
\ea
This definition can be extended to the moments by linearity and the Leibnitz rule and yields `classical Poisson brackets' for the expectation values of basic canonical variables, i.e.\ $\{\langle\hat{q}\rangle,\langle\hat{p}\rangle\}=1$, vanishing Poisson brackets between expectation values and moments $\{\langle\hat{q}\rangle,\Delta(q^{a}p^{b} )\}=0$ and more complicated brackets between the moments \cite{bojski,effcon1}.

In this summary we focus on finite dimensional systems with a single constraint $\hat{C}$ playing the role of the Hamiltonian constraint of general relativity. The Dirac programme requires physical states to satisfy $\hat{C}|\psi\rangle=0$. The spectrum of $\hat{C}$ is essential for the construction of the physical Hilbert space: if zero lies in the discrete part of the spectrum, the physical Hilbert space turns out to be a subspace of the kinematical Hilbert space, while a new Hilbert space with a new (positive--definite) inner product must be constructed for solutions if zero lies in the continuous part of the spectrum. There exist techniques for constructing such physical Hilbert spaces in the latter case \cite{hilbertspace}, however, finding physical Hilbert spaces in practice remains a difficult task.
  
The effective techniques, on the other hand, work for both zero in the discrete or continuous part of the spectrum. At the effective level, physical states must clearly satisfy
\ba\label{qcon}
\langle\hat{C}\rangle(\langle\hat{q}\rangle,\langle\hat{p}\rangle,\Delta(qp),\cdots)=0\q,
\ea
however, this is not sufficient since the expectation value of $\hat{C}$ may be zero even if $\hat{C}|\psi\rangle\neq0$. Furthermore, when solving one (first class) constraint classically we can eliminate an entire canonical pair, while on the quantum phase space after solving (\ref{qcon}) and factoring out its flow we would be left with an infinite tower of (unconstrained) moments of the eliminated canonical pair. Clearly, we must impose a further set of constraints to account for this. It turns out that 
\ba\label{pol}
 C_{\rm pol}:= \langle\widehat{\rm pol}\hat{C}\rangle=0\q,
\ea
where $\widehat{\rm pol}$ stands for all polynomials in the basic canonical operators, provides a correct independent set of constraint functions on quantum phase space which together with (\ref{qcon}) is then equivalent to the Dirac condition \cite{effcon1,effcon2}.\footnote{Note that no Weyl--ordering is imposed on the constraints, for otherwise the constraints fail to be first class \cite{effcon1}. This entails complex--valued quantum constraint functions at the kinematical level, however, physical reality and positivity conditions can be imposed on Dirac observables to describe proper quantum states. In the examples studied thus far, these conditions are preserved by the dynamics \cite{bht2}.}

As a result, we have infinitely many constraints for infinitely many variables. In order to reduce the system to a tractable finite size, we impose a very general semiclassical approximation: assume $\Delta(q^ap^b)=O(\hbar^{(a+b)/2})$ and truncate the system at $\hbar$ by neglecting all terms of higher order. This is consistent with the quantum Poisson bracket structure (\ref{poisson}) which preserves orders in $\hbar$ and generates the flows and dynamics on quantum phase space, once suitable initial data for the relevant expectation values and moments has been chosen.

Rather than defining a quantum theory, the effective approach in its present form provides an effective tool for evaluating the quantum dynamics of given (finite dimensional) systems.

\section{A simple example without a global clock function}

Let us outline the effective approach to semiclassical relational dynamics with an explicit toy model (for a more general discussion and details see \cite{bht1,bht2}). The model we are considering is the isotropic 2D harmonic oscillator with prescribed total energy, whose classical solutions are closed orbits in phase space such that no global clock exists \cite{bht2}. It is subject to the constraint
\ba\label{quant-rov}
\hat{C}|\psi\rangle_{\rm phys}=(\hat{p}_1^2+\hat{p}_2^2+\hat{q}_1^2+\hat{q}_2^2-M)|\psi\rangle_{\rm phys}=0\q.
\ea
The only variables from which we could construct clock functions are the $\hat{q}_i,\hat{p}_i$. In order to obtain a notion of evolution from this {\it a priori} timeless system, we may choose a variable as a local clock, say $q_1$, and deparametrize locally at the classical level by factorizing the constraint 
\ba
C=\left(p_1+H(q_1)\right)\left(p_1-H(q_1)\right), \q\text{where}\q
 H(q_1)=\sqrt{M-q_1^2-p_2^2-q_2^2}\q.\nonumber
 \ea
Quantization of the first factor (in the region where it vanishes) yields a local Schr\"odinger regime
\ba\label{schrod}
i\hbar\frac{\partial}{\partial
q_1}\psi(q_1,q_2)=\hat{H}(\hat{q}_2,\hat{p}_2;
q_1)\psi(q_1,q_2)=\widehat{\sqrt{M-q_1^2-p_2^2-q_2^2}}\,\psi(q_1,q_2)
\ea
with a non--self--adjoint $\hat{H}$ (defined by spectral decomposition) due to non--unitarity in $q_1$ evolution ($q_1$ is not globally valid). Given this construction, we can calculate expectation values and moments in $q_1$ time and compare their evolution to the effective treatment.

At the effective level, on the other hand, we retain 14 kinematical degrees of freedom at order $\hbar$, namely four expectation values $a:=\langle\hat{a}\rangle$, four spreads $\langle(\hat{a}-a)^{2} \rangle_{\rm Weyl}$, and 
 six covariances $\langle(\hat{a}-a)
(\hat{b}-b)\rangle_{\rm Weyl}$, where the $\hat{a},\hat{b}$ are any of the $\hat{q}_i,\hat{p}_i$. At order $\hbar$ the constraint (\ref{quant-rov}) translates---via (\ref{qcon}) and (\ref{pol})---into the following five first class quantum constraint functions \cite{bht2}
\ba\label{effrov}
C&=& p_1^2+p_2^2+q_1^2+q_2^2+(\Delta p_1)^2+(\Delta p_2)^2+(\Delta q_1)^2+(\Delta q_2)^2-M= 0
\nonumber \\
C_{q_{1}}&=& 2p_1\Delta(q_1p_1)+2p_2\Delta(q_1p_2)+2q_1(\Delta q_1)^2+2q_2\Delta(q_1q_2)+i\hbar p_1= 0 \nonumber \\
C_{p_1}&=&
2p_1(\Delta p_1)^2+2p_2\Delta(p_1p_2)+2q_1\Delta(p_1q_1)+2q_2\Delta(p_1q_2)-i\hbar q_1= 0 \nonumber
\\
C_{q_2}&=&2p_1\Delta(p_1q_2)+ 2p_2 \Delta(q_2p_2)+2q_1\Delta(q_1q_2)+2q_2(\Delta q_2)^2+i\hbar p_2 = 0 \nonumber \\
C_{p_2}&=&2p_1\Delta(p_1p_2)+ 2p_2(\Delta p_2)^2+2q_1\Delta(q_1p_2)+2q_2\Delta(q_2p_2)-i\hbar q_2 = 0.\nonumber
\ea
These five constraints generate only four independent flows since at order $\hbar$ we have to deal with a degenerate Poisson structure \cite{bht2}. It is convenient to fix three of these independent flows. Selecting $q_1$ as a relational clock it should not be represented as an operator; we therefore choose a gauge which `projects this clock to a parameter' by setting its fluctuations to zero (fixing three $O(\hbar)$ flows and leaving us with one `Hamiltonian flow')  
\ba\label{zeitgeist}
(\Delta q_1)^2=\Delta(q_1q_2)=\Delta(q_1p_2)=0 \q.
\ea
Indeed, the choice of an internal time variable in the effective framework is best described and interpreted in a corresponding gauge \cite{bht1,bht2}: we refer to such a choice and gauge (e.g.\ (\ref{zeitgeist})) as a {\it Zeitgeist}. After choosing a local relational clock and corresponding Zeitgeist, local relational observables at the effective level are given by correlations of expectation values and moments with the expectation value of the chosen clock (here ${q}_1$) evaluated in its Zeitgeist. We call such state dependent local observables {\it fashionables} because they comprise the complete physical information about the system (at order $\hbar$) as long as the Zeitgeist is valid, but may fall out of fashion when a Zeitgeist necessarily changes at turning points of local clocks.

It turns out that the fashionables in $q_1$ time of the effective framework agree perfectly with those computed in the $q_1$ Schr\"odinger regime (\ref{schrod}) \cite{talk,bht2}. In its turning region, $q_1$ becomes complex--valued and its Zeitgeist incompatible with the semiclassical expansion, such that evolution in $q_1$ breaks down {\it before} the turning point---a signature of non--unitarity. Thus, a new local internal time (here $q_2$) is needed and for a full evolution we must switch between $q_1$ and $q_2$ time. Furthermore, since a given internal time is best described in a corresponding choice of gauge, we must switch also between the $q_1$ and $q_2$ Zeitgeister. Explicit gauge transformations between these Zeitgeister can, indeed, be constructed and initial data can be evolved consistently around the entire closed semiclassical orbit through the turning points of various clock variables \cite{bht2}. It should be emphasized that in each gauge we evolve a {\it different} set of fashionables which highlights the local nature of the relational concept in the absence of global clock functions.

%


\section{Conclusions}

The effective approach summarised here sidesteps the technical problems mentioned in the introduction. In particular, the {\it Hilbert space problem} is avoided altogether since no use of any Hilbert space representation has been made. At the effective level one can make sense of local time evolution and switching between different local clocks  can essentially be achieved by an additional gauge transformation which enables one to handle the {\it multiple choice} and {\it global time problem} and implies the equivalence of different clock choices at semiclassical order. Note, however, that if the relational clock is non--global, it assumes complex values in its turning region \cite{bht1,bht2}. State dependent fashionables arise naturally in this framework which simplifies the {\it observable problem} due to the classical treatment of the effective system. In the example outlined here and other models, the effective evolution agrees with local deparametrizations at a Hilbert space level \cite{bht2}. Finally, the notion of relational evolution disappears in highly quantum states of systems without global clocks \cite{bht1,bht2}. The application of this effective framework to the more interesting closed FRW model with a massive scalar field 
 will appear elsewhere \cite{hkt}.

{\bf Acknowledgments} The author would like to thank Martin Bojowald and Artur Tsobanjan for collaboration on this subject.


\section*{References}
\begin{thebibliography}{9}
\bibitem{talk} slides at \url{http://loops11.iem.csic.es/loops11/index.php?option=com_content&view=article&id=101}
\bibitem{rel} C.~Rovelli, \textit{ Quantum Gravity} (CUP, Cambridge, 2004); C.~Rovelli, Phys.\ Rev.\ {\bf D 42} 2638 (1990), Phys.\ Rev.\ {\bf D 43} 442 (1991); B.~Dittrich, Gen.\ Rel.\ Grav.\ {\bf 39} 1891 (2007), Class.\ Quant.\ Grav.\ {\bf 23} 6155 (2006)
\bibitem{pot} K.~V.~Kucha\v{r}, in \textit{Proc.\ 4th Canadian Conference on General Relativity and Relativistic Astrophysics}, edited by G.~Kunstatter, D.~Vincent and J.~Williams (World Scientific, Singapore, 1992); C.~J.~Isham, in \textit{Integrable Systems, Quantum Groups, and Quantum Field Theories} (Kluwer Academic Publishers, London, 1993);
E.~Anderson ({\it Preprint} arXiv:1009.2157 [gr-qc]); J.~Tambornino ({\it Preprint} arXiv:1109.0740 [gr-qc])

\bibitem{bht1} M.~Bojowald, P.~A.~H\"ohn and A.~Tsobanjan, Class.\
  Quantum Grav.\ {\bf 28} 035006 (2011)
  \bibitem{bht2} M.~Bojowald, P.~A.~H\"ohn and A.~Tsobanjan,  Phys.\ Rev.\  {\bf D 83 } 125023 (2011)  
  \bibitem{bojski} M.~Bojowald and A.~Skirzewski, Rev.\ Math.\ Phys.\ {\bf 18}
713--745 (2006) 

\bibitem{effcon1} M.~Bojowald, B.~Sandh\"ofer, A.~Skirzewski and A.~Tsobanjan, Rev.\ Math.\ Phys.\ {\bf 21} 111 (2009) 

\bibitem{hilbertspace} D.~Marolf ({\it Preprint} gr-qc/9508015); T.~Thiemann, Class.\ Quant.\ Grav.\ {\bf 23} 2211 (2006)

\bibitem{effcon2} M.~Bojowald and A.~Tsobanjan, Phys.\ Rev.\ {\bf D 80} 125008 (2009) 

\bibitem{hkt} P.~A.~H\"ohn, E.~Kubalov\'a and A.~Tsobanjan, {\it to appear}
  
  \end{thebibliography}
\end{document}